        \documentclass{ws-procs9x6}            

\begin{document}

\title{Studies of $\eta^{(')}\pi$ Final States Using GlueX Data}

\author{Colin Gleason, for the GlueX Collaboration}

\address{Department of Physics, Indiana University Bloomington,\\
Bloomington, IN 47405 United States of America\\
$^*$E-mail: colgleas@iu.edu
}

\begin{abstract}

The primary goal of the GlueX experiment at Jefferson Lab is to search for and map the spectrum of light hybrid mesons. 
Many experiments have studied and reported evidence of exotic mesons decaying into $\eta\pi$ and $\eta'\pi$ final states.
With a large acceptance to both charged and neutral particles, GlueX has access to both the neutral, $\gamma p\to\eta^{(')}\pi^{0}p$, and charged, $\gamma p\to\eta^{(')}\pi^{-}\Delta^{++}$, exchanges.
These proceedings will give an overview of the current studies being performed at GlueX in $\eta^{(')}\pi^{-}$ final states.
It will discuss early physics goals and outline the strategy for an amplitude analysis as GlueX begins its quest to illuminate the light hybrid meson spectrum.
\end{abstract}

\bodymatter

\section{Introduction}\label{aba:sec1}
GlueX is a photoproduction experiment located at the Thomas Jefferson National Accelerator Facility in Newport News, VA.
The goal of GlueX is to understand how hadrons are generated from quantum chromodynamics (QCD).
To do this, GlueX will attempt to map the broad spectrum of hybrid mesons, as predicted by lattice QCD\cite{Dudek:2013yja}.
Some hybrids have quantum numbers ($J^{PC}$) that can not be constructed from a $q\bar{q}$ system.
These are known as exotic hybrid mesons.
The search for hybrids at GlueX begins with the lowest lying exotic hybrid, the $\pi_{1}(1600)$.
Eventually, we will work towards measuring the properties of hybrids across their full spectrum which will lead to an understanding of how hadrons arise from QCD.

Final states with $\eta^{(')}\pi$ are an ideal place to search for hybrids due to the fact that exotic $J^{PC}$ arise if an $\eta^{(')}\pi$ system is observed in an odd partial wave.
Many experiments have searched for exotic $J^{PC}$ in $\eta^{(')}\pi$ final--states \cite{GAMS_Alde:1988bv,KEK_Aoyagi:1993kn,VES_Beladidze:1993km, E852_Thompson:1997bs,CB_Abele:1999tf,E852_Adams:2006sa, E852_Ivanov:2001rv, VES_Dorofeev:1999th, VES_Gouz:1992fu, VES_Amelin:2005ry, CLEO_PhysRevD.84.112009, COMPASS_2015303}.
In general, these experiments report a $P$--wave ($J=1$) enhancement corresponding to a $\pi_{1}(1400)$ in $\eta\pi$ final--states and/or a $\pi_{1}(1600)$ in $\eta'\pi$ final--states.
The Joint Physics Analysis Center (JPAC) recently did a coupled--channel fit to the COMPASS data and showed that the $\pi_{1}(1400)$ and $\pi_{1}(1600)$ originate from the same pole, thus resolving the $\pi_{1}$ discrepancy seen in previous experiments \cite{JPAC_Rodas:2018owy, COMPASS_2015303}. 

\section{The GlueX Experiment}
\begin{figure}
	\begin{center}
	\includegraphics[width=4.in]{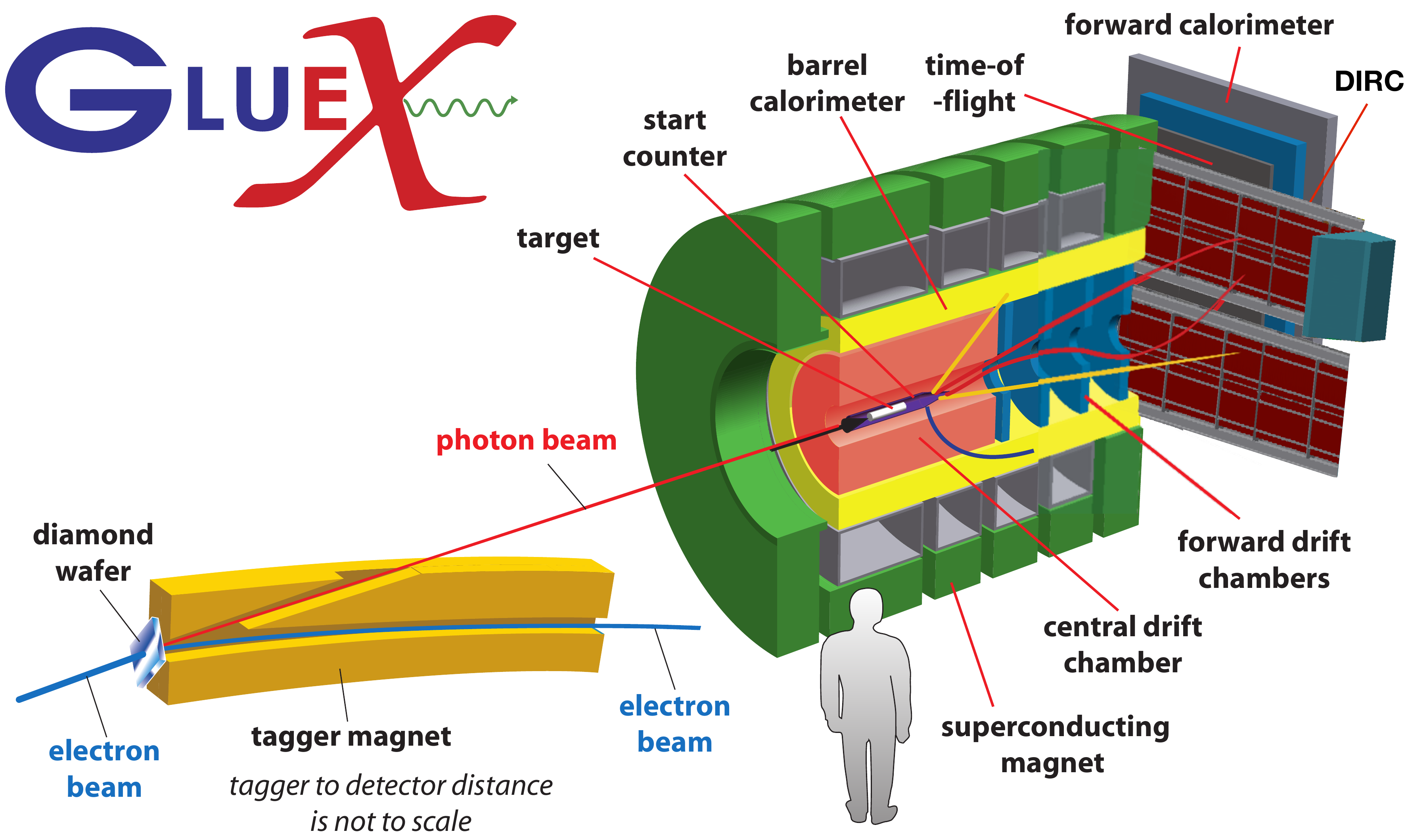}
	\end{center}
	\caption{The GlueX spectrometer.}
	\label{fig:spectrometer}
\end{figure}

A schematic of the GlueX spectrometer can be seen in Figure \ref{fig:spectrometer}.
An electron beam with energies up to 12 GeV is delivered to Hall D and interacts with a thin radiator to produce the photon beam that continues down the beam line towards a LH$_{2}$ target.
The target, central drift chamber, forward drift chamber, and barrel calorimeter are located inside of the solenoid magnet, which operates with a field of 2T.
Further downstream and outside of the solenoid magnet are a time--of--flight detector, forward calorimeter, and a newly installed detection of internally reflected light (DIRC) detector.
GlueX is unique in that it has a real linearly polarized photon beam and nearly full acceptance of both charged and neutral particles.
GlueX began taking data in 2016 and completed Phase--I of its data taking in December of 2018 with $\approx110$ pb$^{-}$ data collected. 
Phase--II begins in December 2019 and will use the newly installed DIRC. 

\section{$\eta^{(')}\pi$ Systems With GlueX}
GlueX has access to multiple channels and decay modes of $\eta^{(')}\pi$ final--states, such as the neutral exchange $\eta^{(')}\pi^{0}p$ and charge exchange $\eta^{(')}\pi^{-}\Delta^{++}$ reactions.
For $\eta\pi$ systems, we are studying $\eta$ decays into $\gamma\gamma$ or $\pi^{+}\pi^{-}\pi^{0}$.
The different decay modes of the $\eta$ should contain the same physics yet have different acceptances and backgrounds.
Additionally, the charge exchange and neutral reactions are complimentary and should provide information on how hybrid mesons are produced.

\subsection{$\gamma p\to\eta\pi^{-}\Delta^{++}$}

\begin{figure}[!htbp]
	\begin{center}
	\includegraphics[width=4.in]{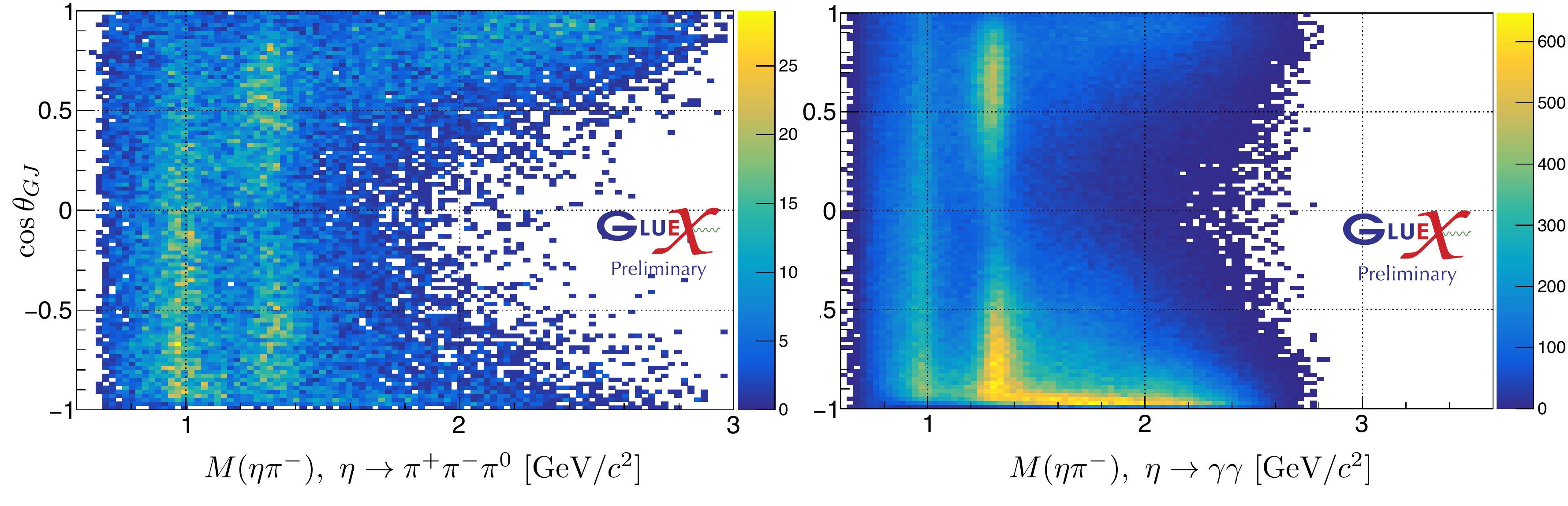}
	\end{center}
	\caption{$\cos\theta_{GJ}$ as a function of $M(\eta\pi^{-})$ where $\eta\to\pi^{+}\pi^{-}\pi^{0}$ (left) and $\eta\to\gamma\gamma$ (right).}
	\label{fig:EtaPiMass}
\end{figure}

Figure \ref{fig:EtaPiMass} shows $\cos\theta_{GJ}$ as a function of $\eta\pi^{-}$ mass for $\eta\to\pi^{+}\pi{-}\pi^{0}$ (left) and $\eta\to\gamma\gamma$ (right), not corrected for acceptance.
In both distributions, clear enhancements are seen in the region of the $a_{0}^{-}(980)$ and $a_{2}^{-}(1320)$, which have the expected $S$ and $D$ wave structures.
One can also see how the acceptances and backgrounds vary between $\eta$ decay modes, particularly at large $\eta\pi$ masses.
There is less acceptance as $\cos\theta\to-1$ for the 3$\pi$ decay when compared to the $2\gamma$ decay.
This leads to a larger amount of background at $\cos\theta\approx-1$ in $\eta\to\gamma\gamma$, which comes from double Regge exchange (e.g. Deck effect).

There are $\approx10$x more statistics in the $\eta\to\gamma\gamma$ than the $\eta\to\pi^{+}\pi^{-}\pi^{0}$ decay mode.
The statistics in the $\eta\to\pi^{+}\pi^{-}\pi^{0}$ channel are comparable to the COMPASS result \cite{COMPASS_2015303}.
A short term physics goal in these channels is to study $a_{0}$ and $a_{2}$ production as a function of momentum transfer $t$.
We hope to use this information to test our understanding of the acceptance and backgrounds.
Additionally, we plan on studying double Regge exchange at large $\eta\pi$ masses.
The goal here would be to constrain this background in the resonance ($a_{0}$, $a_{2}$, $\pi_{1}$) region.

\subsection{$\gamma p\to\eta'\pi^{-}\Delta^{++}$}
Figure \ref{fig:EtaPrimePiMass} shows the $\eta'\pi^{-}$ mass (left) and $\cos\theta_{GJ}$ as a function of $\eta'\pi^{-}$ mass (right).
In the $\eta'\pi^{-}$ mass plot, the red curve is an estimate of the acceptance, the black points are data in the region of the $\eta'$, and the blue points are from the $\eta'$ sidebands.
The $\cos\theta_{GJ}$ distribution is neither sideband subtracted nor corrected for accpetance.
\begin{figure}[!tbh]
	\begin{center}
	\includegraphics[width=4.in]{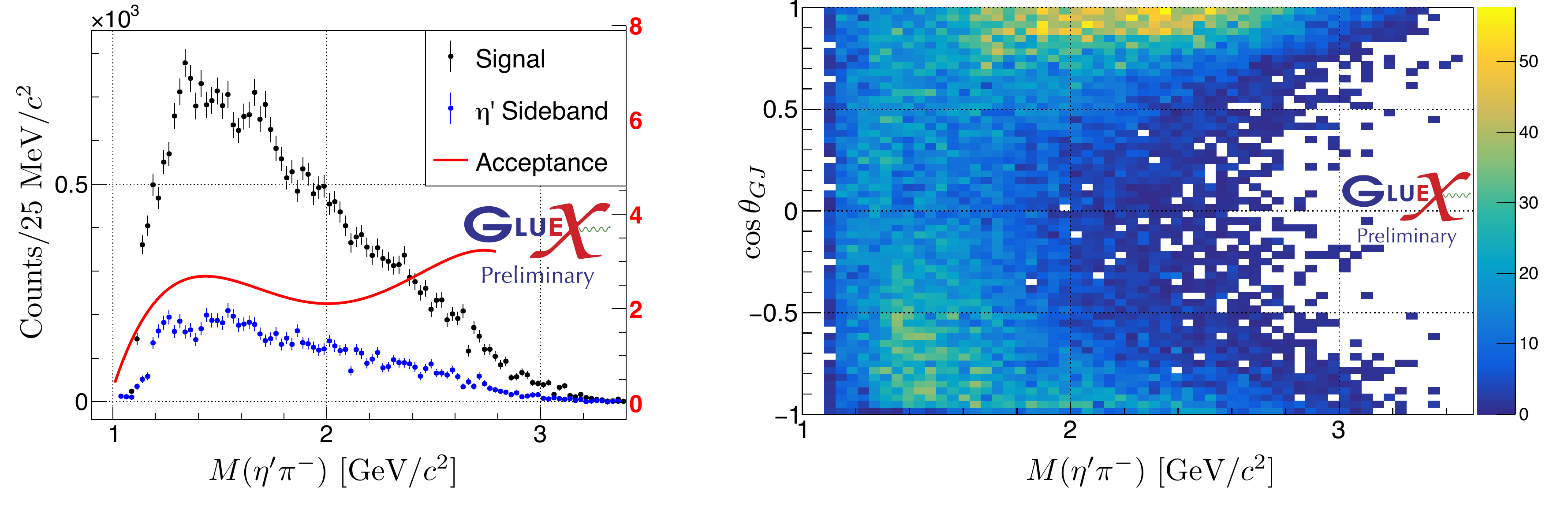}
	\end{center}
	\caption{Left: Invariant mass of the $\eta'\pi^{-}$ system for the signal region (black), sideband regions (blue), and acceptance (red). Right: $\cos\theta_{GJ}$ as a function of $\eta\pi$ mass.}
	\label{fig:EtaPrimePiMass}
\label{fig:EtaPrimePiMass}
\end{figure}

Based of the previous searches for the $\pi_{1}(1600)$, we would expect an enhancement in the $\eta'\pi^{-}$ mass around 1600 MeV, which can be seen in our $\eta'\pi$ mass distribution.
If there is $\pi_{1}\to\eta'\pi^{-}$, we would expect to see a $P$--wave interfering with an $S,~D,$ or $G$--wave in the angular distribution.
At around 1300 MeV, we see a $D$ wave structure corresponding to the $a_{2}(1320)$.
As one moves towards higher $\eta'\pi$ masses, an asymmetry between backward and forward begins to stand out.
There is a broad enhancement at $\approx$1400-1700 MeV at backward angles that is not apparent at forward angles.
Above $\approx1800$ MeV and at $\cos\theta_{GJ}\approx1$, we have backgrounds that will need to be subtracted.
Even though we see some interesting features in the angular distributions, we can not say anything concrete in regards to hybrids until an amplitude analysis is performed.

\section{Summary}

GlueX is beginning its search for hybrid mesons by studying $\eta^{(')}\pi$ final--states.
The eventual goal of this work is to confirm the $\pi_{1}$ pole position extracted by JPAC with different backgrounds and production mechanism than COMPASS.
This work presented mass and angular distributions for the reactions $\gamma p\to\eta\pi^{-}\Delta^{++}$ and $\gamma p\to\eta'\pi^{-}\Delta^{++}$.
We are currently developing the tools and procedures necessary to perform an amplitude analysis on these channels.

\bibliographystyle{ws-procs9x6} 
\bibliography{mybib}
\end{document}